\documentclass[twocolumn,superscriptaddress]{revtex4}
\usepackage{graphicx}
\usepackage{amsmath,amssymb}

\begin{document}
\title{Distributed self-regulation of living tissue. Effects of nonideality}

\date{\today}

\author{Wassily Lubashevsky}
 \email{kloom@mail.ru}
 \affiliation{Moscow Technical University of Radioengineering, Electronics, and Automation,
    Vernadsky 78, 119454, Moscow Russia}
\author{Ihor Lubashevsky}
    \email{ialub@fpl.gpi.ru}
    \affiliation{A.M. Prokhorov General Physics Institute, Russian
    Academy of Sciences, Vavilov Str. 38, 119991 Moscow, Russia}
    \affiliation{Moscow Technical University of Radioengineering, Electronics, and Automation,
    Vernadsky 78, 119454, Moscow Russia}
\author{Reinhard Mahnke}
\email{reinhard.mahnke@uni-rostock.de}
\affiliation{Universit\"at Rostock, Institut f\"ur Physik, 18051 Rostock, Germany}

\begin{abstract}

Self-regulation of living tissue as an example of self-organization phenomena in active fractal systems of biological, ecological, and social nature is under consideration. The characteristic feature of these systems is the absence of any governing center and, thereby,  their self-regulation is based on a cooperative interaction of all the elements. The paper develops a mathematical theory of a vascular network response to local effects on scales of individual units of peripheral circulation.

First, it formulates a model for the self-processing of information about the cellular tissue state and cooperative interaction of blood vessels governing redistribution of blood flow over the vascular network. Mass conservation (conservation of blood flow as well as transported biochemical compounds) plays the key role in implementing these processes. The vascular network is considered to be of the tree form and the blood vessels are assumed to respond individually to an activator in blood flowing though them.

Second, the constructed governing equations are analyzed numerically. It is shown that at the first approximation the blood perfusion rate depends locally on the activator concentration in the cellular tissue, which is due to the hierarchical structure of the vascular network. Then the distinction between the reaction threshold of individual vessels and that of the vascular network as a whole is demonstrated. In addition, the nonlocal component of the dependence of the blood perfusion rate on the activator concentration is found to change its form as the activator concentration increases.
\end{abstract}

\maketitle

\section{Distributed self-regulation in active hierarchical systems}

For a wide class of biological, ecological and social systems, including economic ones a generalized notion of ``homeostasis'' can be introduced. By this term we understand the internal conditions required for such a system to survive as well as its ability to maintain them. Living multicellular organisms can exist if only the temperature, oxygen concentration, etc. are in certain intervals, giving rise to a large number of mechanisms controlling these conditions (for introduction see, e.g., \cite{HHom}). Ecological communities comprising many species continue to exist under environmental perturbations due to complex ``predator-prey'' relationships maintaining their structure. In particular, in microbial communities predator microorganisms act as homeostatic regulators to correct microbial imbalances and restor the balanced environment \cite{EHom1}, density-dependent migration sustains the populations of birds, fishes, and insects \cite{EHom2}, complex species interaction was experimentally revealed in the dynamics of ecosystems of several deserts in Arizona \cite{EHom3}. There are a variety of social and economic systems such as large firms and enterprizes, product and service markets whose dynamics is governed by self-regulation. In these systems the behavioral standards, social norms, can be regarded also as the homeostasis characteristics (see, e.g., \cite{SHom1,SHom2,SHom3}).

When one of the homeostasis parameters deviates from its normal value and comes close to the destruction threshold the system has to respond to this event in order to prevent a further variation of this parameter. For a system with many elements its response to such perturbations is usually implemented via reaction of a special subsystem governing the functioning of the system as a whole. In living tissue blood flow through the vessel network transporting various biochemical compounds governs the homeostasis of the cellular tissue. In a relatively large product market agents of various levels form a trade network and perturbations in the supply-demand equilibrium are damped, in principle, via the reaction of this trade network as a whole. We will call these subsystems life-support networks.

A life-support network is to be of hierarchical structure, because it not only controls the system homeostasis but also supplies the elements with the required ``nutrients'' and withdraws their ``life activity'' products. Typically ``nutrients'' come into a system centrally from the external environment, so, before getting every element their flow branches many times.

\begin{figure}
\begin{center}
\includegraphics[width = 50mm]{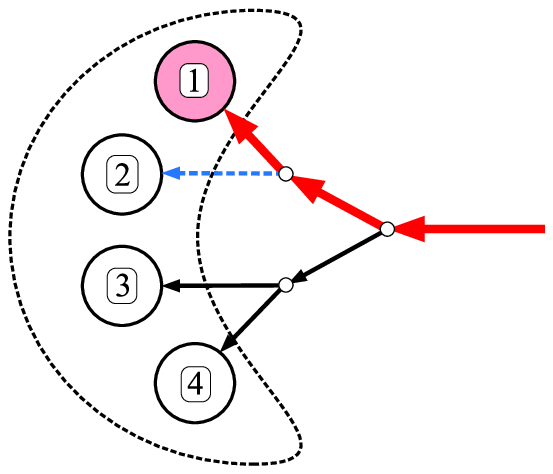}
\end{center}
\caption{(Color online) Illustration of the interference in the element functioning caused by the hierarchical organization of the life-support network; intensification of element 1 activity requiring the ``nutrient'' flow to be increased disturbs the ``nutrient'' flow at elements 2.}
\label{Fig1}
\end{figure}

In the ideal case a life-support network has to operate in such a manner that every element be supplied with the amount of ``nutrients'' required for its current activity. However, for the product flow to very in a given branch of the terminal hierarchy level it has to change at all the levels. This flow redistribution in turn can induce the flow perturbations in the other terminal branches, neighboring and distant ones, thereby disturb the functioning of the corresponding elements and cause an erratic behavior of the system as a whole (Fig.~\ref{Fig1}).

The existence of tremendous amount of natural systems with hierarchical organization enables us to suppose that this destructive interference should be suppressed or, at least, depressed by the appropriate self-regulation processes. If there is no special unit governing the functioning of such system as a whole the self-regulation can be implemented only via a cooperative interaction between all its elements. The latter feature allows us to call such regulation processes distributed self-regulation.

Up to now the specific mechanisms by which the distributed self-regulation arises are far from being understood well. In order to construct an appropriate model for the distributed self-regulation it is necessary to elucidate two aspects. The first one is the self-processing of information. In a system without a governing center none of the  elements forming the life-support network possesses the complete information about the system state. Moreover, it really needs only a piece of information required for the individual functioning. The second aspect concerns the cooperative mechanism governing the product flow redistribution over the life-support network. None of these elements controls individually the product flow even through itself; responding to the appropriate information it can change only its own characteristics affecting the product flow. So the required redistribution of the product flow must be a cumulative effect of their individual contributions.

Let us note some natural systems with the distributed self-regulation. First, it is living tissue, the main object of the present analysis, which will be discussed in detail in Sec.~\ref{PBack}.
Second, hierarchical organization is encountered frequently in ecological communities based on trophic food pyramids \cite{ONeil} (see also a review~\cite{Platt}). Their levels are made up of animals comparable in size and playing much the same role in the prey-predator relationships. The linkages from small organisms generally vary over smaller scales. Larger animals that dominate these smaller organisms do so over larger scales. In such systems variations in species population disturbs locally the prey-predator equilibrium and the induced self-regulation processes should sustain biodiversity \cite{BD1,BD2}. This concept is applicable also to socio-ecological systems comprising ecological communities, landscapes, and human activities (see, e.g., \cite{SES}).

Third example is social networks describing many social, ecological, and technological systems (for a review see \cite{SN1,SN2,SN3}). There are two categories of social networks \cite{OS}. One comprises the self-organized networks, for instance, acquaintance webs or e-mail networks, where hierarchy structures develop and evolve due to local agent interactions with the lack of global goals (see, e.g., \cite{Nature1} and references therein as well \cite{HomSys1,HomSys2}).
The other category is relevant to social systems with human collective efforts towards a given objective, e.g., large companies, distributed technology development, financial networks, and commodity markets or markets based on a certain raw material. The open-source communities \cite{OS} as well as problem-solving networks aimed at product development \cite{Engin} also belong to the second category. The interplay between bottom-up decision making periphery aimed at the overall goals of the community and top-down driven hierarchical organization can be implemented via response of such groups changing, for example, the number of programmers involved in their activity. A similar situation can be met in financial webs or market networks (see, e.g. \cite{SHom1,Bankers}). Hierarchical networks of administrative structures should have the capability of rearranging themselves to respond properly to failures, crises and disasters \cite{Helbing}.

Papers~\cite{we1,we2,we3,we4,we5} proposed several models for self-regulation of living tissue, a goods market based on a certain raw material, and ecological community. Their common feature is, first, the hierarchial self-processing of the information about the system state that is implemented via, for instance, conservation of blood flow and transported biochemical compounds or conservation of the raw material and money flow at the market network. Second, it is the individual response of, for example, blood vessels to  biochemical compounds in blood flow going trough them or firms to disturbance in the local supply-demand equilibrium. It has been found that there exist special conditions under which self-regulation of such systems is perfect. Namely, if the functioning of an element has changed and it needs an increased amount of ``nutrients'' then the life-support network is able to do this in such a manner that the ``nutrients'' flow at the other elements be not disturbed at all.

Unfortunately, the perfect self-regulation requires an ideal behavior of elements, so in real systems it cannot take place rigorously and should be regarded as the first approximation only. Since studying such systems at the initial stage it is worthwhile to single out some characteristic examples of these objects and to investigate them individually. The present paper considers living tissue in this way and analyze its response to local effects when the vessel properties are not ideal.

\section{Physiological background and fundamentals of distributed self-regulation}\label{PBack}

To justify the key points of the model to be developed let us note the relevant properties of the human physiology referring mainly textbook~\cite{LTReg} with respect to the general features.

\subsection*{The architectonics of vessel network}

\begin{figure}
\begin{center}
\includegraphics[width=65mm]{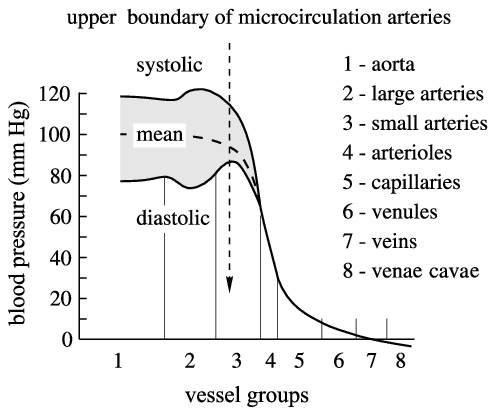}
\end{center}
\caption{Schematic illustration of the blood pressure distribution over the vessels of systemic circulation and the roles that various groups of vessels play in blood flow distribution (after \cite{Kositsky} with some modification).}
\label{Fig2}
\end{figure}

When an organ of human body is active and its tissue needs greatly increased supply of nutrients, blood flow can locally grow by 20 or 30 times. In contrast, the heart normally cannot increase its cardiac output more than 4 to 7 times greater than when at rest. So, dilating or constricting, the blood vessels of each organ have to control \emph{locally}  blood flow at the required level.

Figure~\ref{Fig2} exhibits the distribution of blood pressure over the vessels of systemic circulation, which can be classified into ``conveying'' and ``delivering'' types \cite{Zamir}. The large vessels, making up the ``conveying'' type, maintain a constant value of the mean blood pressure inside the large arteries and damp pressure pulsations via the special mechanism based on the baroreceptors located in the artery walls. As a result, at the entrance to the ``delivering'' circulation system of individual organs the blood pressure can be considered as being fixed at the first approximation if the whole organism is not under extreme conditions.

Below the circulation system of an organ comprising  small arteries and veins, arterioles and venules together with capillary network will be referred to as a microcirculatory bed. As seen in Fig.~\ref{Fig2} the main decrease in the blood pressure falls on the small arteries and arterioles and the blood pressure drop is rather uniformly distributed over many vessels of different lengths. Thereby the net resistance of a microcirculatory bed to blood flow is determined by all its arteries and arterioles rather than vessels of one fixed length.

In general, each artery entering an organ branches six to eight times before the arteries become small enough to be called arterioles, then the arterioles themselves branch two to five times before supplying blood to the capillaries. The arteries and arterioles are highly muscular and their lumen diameters can change manyfold. The branching of these ``delivering'' vessels is rather symmetric; the daughter arteries of a mother artery are typically of the same size and form similar angles with the the parent vessel.

\begin{figure}
\begin{center}
\includegraphics[width=80mm]{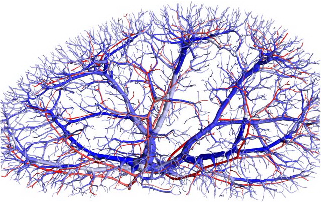}
\end{center}
\caption{(Color online) Illustration of the vascular network layout made up of the counter-current pairs. The figure exhibits the rat renal vasculature reconstructed from computer tomography images (after \cite{ratvessels}, used with permission).}
\label{Fig3}
\end{figure}

Microcirculatory beds are typically arranged in a space-filling manner such that the terminal arterioles and venules be distributed in the tissue uniformly. Besides, rather often the layout of their arterial and venous networks is mainly made up of the counter-current pairs consisting individually of an artery and vein located in space in close proximity to each other. Figure~\ref{Fig3} illustrates this counter-current vessel arrangement. Possible mechanisms of such pattern formation are discussed, in particular, in \cite{CCur}.

\subsection*{The mechanisms of the tissue self-regulation}

To illustrate the state of art we discusses the tissue self-regulation mainly in the brain. In this organ the homeostasis should be controlled by blood flow most effectively and its mechanisms have attracted much attention during the last decade (for a review see \cite{NRRev1,NRRev2,NRRev3,NRRev4}).

There is a lot of evidence that regulation of cerebral circulation is highly localized and cannot be governed \emph{solely} by the products of tissue metabolism (see a review~\cite{NR0} and references therein). For example, during local activation of a small region of the brain the corresponding increase in the blood perfusion rate is caused by the dilation of not only small arterioles in it but also relatively large pial arteries that supply directly the activated region with blood~\cite{NR2,NR3,NR4}. On one hand, exactly these arteries offer the greatest resistance to flow and, consequently, are the main site of flow control~\cite{NR6}. On the other hand, the dilated pial arteries lay outside the activated region and, thus, cannot be affected directly by biochemical products of cell activity in it. Thereby other mechanisms should also contribute to the blood flow regulation. The main hypothesis about these mechanisms noted in many textbooks (e.g., \cite{LTReg}) is based on the release of vasodilator substances such as nitric oxide (NO) by the cells of artery walls. Rapid blood flow through the arteries and arterioles causes shear stress on their cells, inducing the release of NO and finally the vessel dilation. The upstream propagation of this dilation is assumed to be conducted by the cell-to-cell interaction (through homocellular gap junctions with the ions Ca$^{2+}$ playing the key role) or just the effect of the successive blood pressure redistribution (see, e.g., \cite{NRRev1}). Yet these models are rather far from being verified well in vivo \cite{NRRev4}, and other mechanisms are under discussion (see, in particular, \cite{No1,No2,No3,No4,No5}). Moreover, even the fact of observing in vivo upstream coupled waves of Ca$^{2+}$ and arteriolar dilation \cite{VivoWaves} poses a question as to how these waves propagating in both the directions along the arteries pass the points of vessel branching and transfer the information about the cellular tissue states.

The upstream artery delation can be alternatively caused by artery-vein interaction; for a review see \cite{AVInt1} and, in particular, paper~\cite{AVInt2} comparing several possible mechanisms with each other as well as work~\cite{AVInt3} studying this phenomenon in the brain. The main idea is that the downstream transport of biochemical substance with blood flow going through the vein network gives rise to an affective upstream processing of information about the cellular tissue state. The exchange of biochemical compounds between arteries and veins forming counter-current vessel pairs was assumed to arise via diffusion through the surrounding cellular tissue, being justified theoretically \cite{Pop1,Pop2}.

Neural stimuli should also contribute to the \emph{local} regulation of blood flow. In the brain the neural mechanisms of the blood flow regulation are due to the neurovascular units \cite{NVU1,NVU2,NVU3}. Roughly speaking, the neurovascular unit consists of vessels of microcirculation, interneurons, and astrocytes. Interneurons give rise to short pathways between between the tissue cells in the activated region and vessels supplying it with blood. Astrocytes together with interneurons affect the muscles of vessel wall causing dilation or contraction (for details see \cite{Mech} and references therein).

\begin{figure}
\begin{center}
\includegraphics[width=85mm]{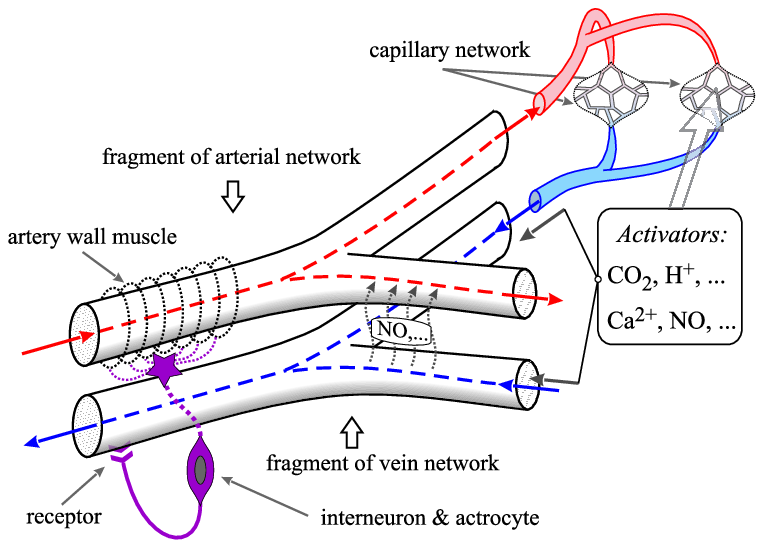}
\end{center}
\caption{(Color online) Schematic illustration of physiological processes used in contracting the developed model.}
\label{Fig4}
\end{figure}

The model to be developed actually combines the counter-current artery-vein interaction and the neural mechanism of local blood flow regulation (Fig.~\ref{Fig4}). It assumes that the products of cell activity, for example, CO$_{2}$, H$^{+}$, and bioactive substances, including, the ions Ca$^{2+}$, NO, etc. are accumulated in the cellular tissue and then with blood enter the vein bed. Below these biochemical substances are called activators. Receptors embedded into the vein walls detect the concentration $\Theta$ of the activators in blood flow going through the corresponding veins. Then the ``reading'' of the receptors are transferred by the neural mechanism to the arteries coupled with these  veins, which determines the degree of the artery dilation or contraction. It should be noted that the substance diffusion between the nearest artery and vein can also imitate the action of these receptors.

The next subsection demonstrates that the proposed model does give rise to the required upstream response of the arteries to disturbance in the cellular tissue homeostasis. Moreover, if this artery response could be caused by the upstream propagation of some biochemical substance along the arteries that obeys conservation at the branching points of the artery tree then the developed model could also allow for this mechanism after minor modification.

\subsection*{The fundamentals of the distributed self-regulation}

\begin{figure}
\begin{center}
\includegraphics[width=80mm]{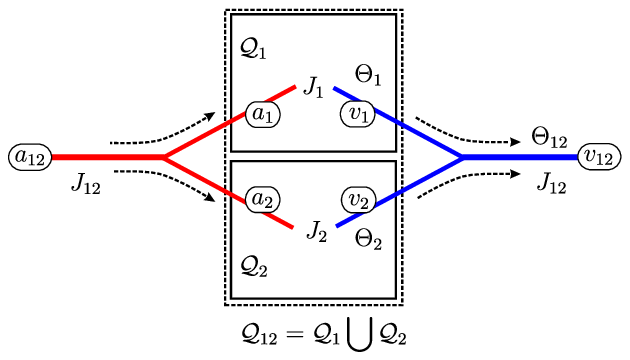}
\end{center}
\caption{(Color online) Illustration of the information self-processing caused by mass conservation in the blood flow through the venous bed.}
\label{Fig5}
\end{figure}

Figure~\ref{Fig5} illustrates the implementation of the self-processing of information about the states of living tissue. Let the activators be uniformly distributed in the domains $\mathcal{Q}_1$ and $\mathcal{Q}_2$ individually with concentrations $\Theta_1$ and $\Theta_2$, respectively. So the values $\Theta_1$ and $\Theta_2$ characterize the states of the living tissue in these domains and should determine the response of arteries $a_1$ and $a_2$ directly supplying the living tissue domains $\mathcal{Q}_1$ and $\mathcal{Q}_2$ with blood, oxygen, and nutrients. Draining the given domains via the capillary network being in local quasiequilibrium with the surrounding tissue gives rise to the same values of the activator concentrations in veins $v_1$ and $v_2$ forming the counter-current pairs with the arteries $a_1$ and $a_2$. The mass conservation at the shown branching points reads
\begin{eqnarray}
\label{2.1a}
  J_{12} &=& J_1 + J_2\,, \\
\label{2.1b}
  J_{12}\Theta_{12} &=& J_1\Theta_{1} + J_2\Theta_{2}\,,
\end{eqnarray}
where $J_1$, $J_2$, and $J_{12}$ are the blood flow rates in these vessels. Whence it follows that the concentration of activators in the parent vein $v_{12}$ equal to
\begin{equation}\label{2.2}
  \Theta_{12} = \frac{J_1\Theta_{1} + J_2\Theta_{2}}{J_1 + J_2}\,.
\end{equation}
is the mean values of the activator concentration in the joint domain $\mathcal{Q}_{12}=\mathcal{Q}_1\bigcup\mathcal{Q}_2$ supplied with blood directly by the pair of the artery $a_{12}$ and the vein $v_{12}$. In this averaging the individual blood flow rates $J_1$, $J_2$ play role of the weight coefficients. Thereby, the activator concentration $\Theta_{12}$ is exactly the quantity characterizing the state of the living tissue domain $\mathcal{Q}_{12}$ as whole and required for the parent artery $a_{12}$ to respond properly to variations in the state of living tissue in $\mathcal{Q}_{12}$.

Generalizing this analysis and assuming the mass conservation to hold at all the branching points of the vein bed we can write
\begin{equation}\label{2.3}
    \Theta_v = \frac{1}{J_v} \sum_{v_c\in v} \Theta_{v_c}J_{v_c}\,,\quad J_v = \sum_{v_c\in v}J_{v_c}\,,
\end{equation}
where $\Theta_v$ and $J_v$ are the concentration of activators and the blood flow rate at a given vein $v$ and the sum runs over all the venules connected directly with the capillary network and originating from the vein $v$. For the sake of simplicity, the terms ``arteriole'' and ``venule'' will be used to refer to such smallest arteries and veins related connected directly with the capillary network, whereas the terms ``artery'' and ``vein'' will be corresponded to the other vessels of arterial and venous beds. Let $\mathcal{Q}_{v_c}$ be the tissue region supplied with blood via the coupled arteriole and venule as a whole. All such domains will be called the elementary domains of living tissue. Then expression~\eqref{2.3} gives us the mean value of the activator concentration in the domain $\mathcal{Q}_v = \sum_{v_c\in v}\mathcal{Q}_{v_c}$ composed of all the elementary domains whose venules originate from the vein $v$.

The receptors located in the vein walls detect the presence of activators in the blood flow going through the corresponding veins by measuring their concentration $\Theta_v$. Via the assumed communication between the artery and vein of a counter-current pair of vessels the receptor signal is transmitted to the artery, causing its dilation or contraction to the degree determined by the value $\Theta_v$.  The hither the concentration, the stronger the artery dilation should be. The effect of artery reaction will be measured in the dependence of its hydrodynamic resistance $R$ on the activator concentration $\Theta$ in the coupled vein.

\begin{figure}
\begin{center}
\includegraphics[width=75mm]{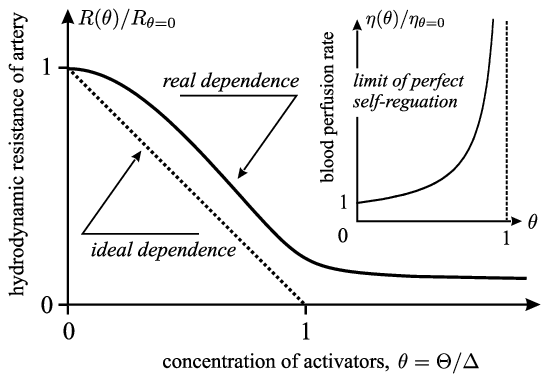}
\end{center}
\caption{The resistance of an artery to blood flow through it vs the concentration of activators in the coupled vein.
The model of perfect self-regulation assumes this dependence to be of the ideal form and the resulting dependence of the blood perfusion rate on the local value of activator concentration in the cellular tissue is shown in inset.}
\label{Fig6}
\end{figure}

When one of the homeostasis parameters comes close to the boundary of the tolerance zone the tissue cells should generate an increased amount of the activators to signalize the system about the critical state, causing the blood flow rate to grow. For example, when the tissue temperature deviates from the normal value 36.6$^\text{o}$~C and gets the boundary about 44--45$^\text{o}$~C the blood flow rate has to increase essentially to prevent the further temperature growth. In the superficial tissue such temperature growth can force the blood flow rate to increase tenfold or higher \cite{SFG}. By the symbol $\Delta$ let us denote the concentration of activators corresponding to such critical situations. In other words, $\Delta$ specifies the threshold in the individual artery response; when the activator concentration in the coupled vein attains the value $\Delta$,  the artery has to dilate maximally. The general form of the adopted dependence of the artery resistance $R$ to blood flow through it on the activator concentration $\Theta$ in the coupled vein is depicted in Fig.~\ref{Fig6} by solid line. Below as well as in Fig.~\ref{Fig6} the activator concentration is measured in the dimensionless units, $\theta = \Theta/\Delta$.

\begin{figure}
\begin{center}
\includegraphics[width=85mm]{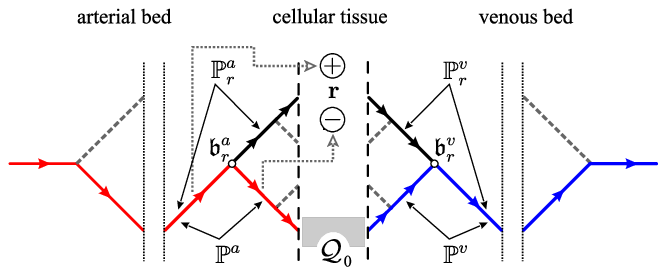}
\end{center}
\caption{(Color online) Schematic illustration of the vessel interaction giving rise the perfect self-regulation of living tissue when the $R(\theta)$-dependence is of the ideal form.}
\label{Fig7}
\end{figure}

When the $R(\theta)$-dependence is linear and can get zero value at $\theta=1$ the vascular network is characterized by the perfect self-regulation~\cite{we1,we2,we3}. Let us define the blood perfusion rate $\eta$ at a point $\mathbf{r}$ of the living tissue as the ratio
\begin{equation}\label{eta}
    \eta(\mathbf{r}) := \frac{J_{v_c}}{V_{v_c}}\,\quad\text{for}\quad \mathcal{Q}_{v_c}\ni\mathbf{r}\,,
\end{equation}
where $V_{v_c}$ is the volume of the elementary domain $\mathcal{Q}_{v_c}$ containing the point $\mathbf{r}$. In these terms the perfectness of the vascular network response is reduced to the statement that the blood perfusion rate $\eta(\mathbf{r})$ depends \emph{only} on the activator concentration $\theta(\mathbf{r})$ at the same point of cellular tissue. It is the result of cooperative interaction of all the arteries in governing the blood flow redistribution over the vascular network, which is illustrated in Fig.~\ref{Fig7}. To explain this fact let us assume that cellular tissue in the elementary domain $\mathcal{Q}_0$ comes close to the tolerance zone and, as a result, the concentration of activators in it increases essentially. Then the activator concentration in all the veins forming up the path $\mathbb{P}^v$ also grows, causing the dilation of all the arteries belong to the path $\mathbb{P}^a$. To analyze the effect of this dilation beyond the domain $\mathcal{Q}_0$ we consider the induced variation of the blood perfusion rate at a point $\mathbf{r}\notin \mathcal{Q}_0$. The widening of all the arteries located on the path $\mathbb{P}^a$ before the branching point $\mathfrak{b}_\mathbf{r}^a$ (i.e. between the entrance to the root artery and this branching point) contributes individually to the increase of the blood perfusion rate at the point $\mathbf{r}$ because their dilation increases the total blood flow through the given vascular network. This effect is denoted by the sign ``+'' in Fig.~\ref{Fig7}.  In contrast, the widening of the arteries located after the branching point $\mathfrak{b}_\mathbf{r}^a$ have to decrease partially the blood perfusion rate $\eta(\mathbf{r})$ because their dilation gives rise to a certain dominance of these arteries in the blood flow redistribution after passing the branching point $\mathfrak{b}_\mathbf{r}^a$. This effect is designated by the sign ``-'' in  Fig.~\ref{Fig7}. As proved in works~\cite{we2,we3} all these contributions of the arteries compensate each other keeping up the value of the blood perfusion rate unchanged at the distant points. Naturally deviation of the $R(\theta)$-dependence from the ideal form breaks this compensation. In this case the blood perfusion rate will depend on the activator concentration also at distant points of cellular tissue.

Finalizing this subsection we also note the fact that in the case of perfect self-regulation the threshold $\Delta$ of the individual vessel response and a similar value $\Theta_\text{th}$ characterizing the response of the vascular network as a whole to increase in the activator concentration in cellular tissue coincide with each other. Indeed, the quantity $\Theta_\text{th}$ can be regarded as such a value of the activator concentration in cellular tissue that gives rise to a substantial increase in the blood flow resistance $Z$ of the vascular network as a whole and, as result, to significant growth of the perfusion rate. In the case of the perfect self-regulation due to the feasibility for the vessel resistance to get zero value the blood perfusion rate $\eta$ depends on $\theta$ as
\begin{equation*}
 \eta(\theta) = \frac{\eta_{\theta=0}}{1-\theta}\,,
\end{equation*}
which allows us to set $\Theta_\text{th} = \Delta$. In the real situation such a singularity cannot arise and the inequality $\Theta_\text{th} > \Delta$ should hold.

\section{Model of living tissue}

The model for the living tissue response to local perturbations consists of three units for the vascular network architectonics, the regularities governing blood flow redistribution, and the description of individual vessel behavior.

\subsection{Vascular network architectonics}

The model for the vascular network architectonics assumes the following.

First, the arterial and venous beds are individually of the tree form \footnote{There are special vessels called anastomoses that connect, e.g., arteries belonging to different vessel branches, endowing vascular networks with a more complex topology. Nevertheless in many organs their relative number is rather low and at the first approximation the presence of anastomoses is of minor importance. The case of high number of anastomoses as it is in the brain requires an individual investigation.}. Second, the branching of vessels is symmetrical, i.e. at the branching points of the arterial bed the radii and lengths of daughter arteries are the same as well as are the angles formed by the daughter arteries with their mother artery. A similar statement holds also for the points of the vein merging. Third, the vessels of each level are distributed uniformly in the living tissue domain. This domain is considered to be a square or cub \footnote{The chosen shapes of the living tissue domain mimic superficial and volumetric microcirculatory beds having similar dimensions in all the directions. The vascular network of organs with essential asymmetry requires also an individual consideration because it should contain ``conveying'' and ``delivering'' parts playing different roles in the blood flow redistribution inside the organ.}.

\begin{figure}
\begin{center}
\includegraphics[width=75mm]{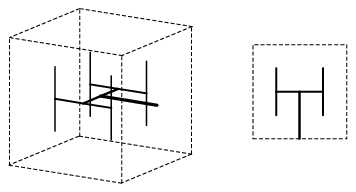}
\end{center}
\caption{Dichotomic model of the vessel tree and its characteristic fragments of the embedding into 2D and 3D domains.}
\label{Fig8}
\end{figure}

The first and second assumptions enable us to introduce the vessel hierarchy and to order all the arteries and veins according to their position in the hierarchical structure. The arteries and veins reachable from the root artery or vein by passing the same number of the branching points (without return) belong to one level of the hierarchy. All the arteries or veins of one level are identical in properties. In what follows the root vessel will be labeled with index $n=0$, the vessels of the next level are labeled with index $n = 1$, and so on. The third assumptions allows us to specify the vessel arrangement in the living tissue domain using a self-similar embedding of the vessels. For example, the length of daughter arteries $l_{n+1}$ is related to the length $l_n$ of the mother artery by the ratio $l_n/l_{n+1} = f$ taking the same value $f$ for all the levels. Rigorously speaking, this relationship holds only for the transitions between two or three successive levels for 2D and 3D models. However, in order to not overload description we will use this statement where it does not lead to misunderstanding. In particular, for the 2D and 3D cases we set $f_{2D} = \sqrt{2}$ and $f_{3D}=\sqrt[3]{2}$, respectively. Figure~\ref{Fig8} depicts the characteristic fragment of the vascular network that illustrates the embedding of vessels into the living tissue domain adopted in the present model. The total number of the hierarchy levels (not counting the root vessels) is divisible by 2 or 3 in 2D or 3D cases.

\begin{figure}
\begin{center}
\includegraphics[width=\columnwidth]{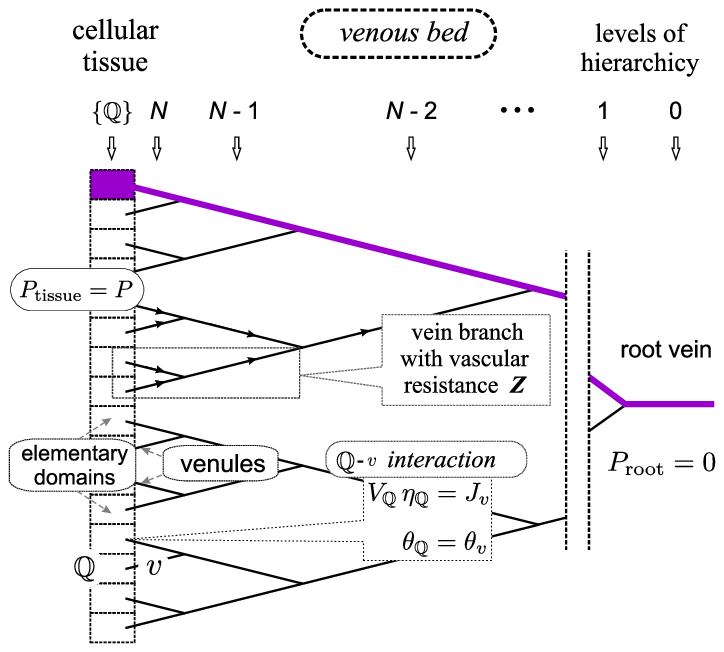}
\end{center}
\caption{(Color online) Vascular network model reduced to the venous bed and its reaction to the distribution of activators in the cellular tissue. Here, in particular, $\mathcal{Q} = \mathcal{Q}_{v_c}$ designates the elementary domain related to the venule $v= v_c$.}
\label{Fig9}
\end{figure}

Besides, for the sake of simplicity, the arterial and venous beds are assumed to be the mirror images of each other with respect to all the properties, including the vessel response. This artificial assumption just simplifies the description of the vascular network response to perturbations in the tissue homeostasis. Indeed, the vascular resistance of the real venous beds is small in comparison with that of the arterial beds. Therefore for a vascular network made up of the counter-current artery-vein pairs the blood flow patterns of the arterial and venous beds should be the mirror images of each other at the first approximation. So the latter assumption simply mimics this situation and enables us to speak only about veins and their dilation or contraction in the response to variations in the activator concentration in blood flow through a given vein. Naturally, in order to get the right values of the blood flow rates in this model we have to increase the blood pressure drop across the vascular network twice or, what is the same, to fix formally the pressure inside the cellular tissue equal to the initial pressure at the entrance to the root artery. Figure~\ref{Fig9} illustrates the adopted model for the venous bed as well as the mechanism of its response to the variations in the activator concentration in the cellular tissue.

\subsection{Governing equations for blood flow distribution}

\begin{figure}
\begin{center}
\includegraphics[width=60mm]{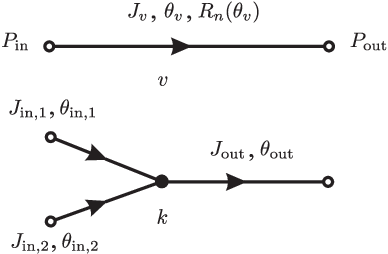}
\end{center}
\caption{The main units of the mathematical description of blood flow distribution over the vein bed.}
\label{Fig10}
\end{figure}

Blood flow distribution over the vein network $\{v\}$ (Fig.~\ref{Fig9}) is described as follows. First, to blood flow with transported activators through every vein $v$ we ascribe the blood flow rate $J_v$ and the activator concentration $\theta_v$. The state of this vein is quantified in terms of the vessel resistance $R_n(\theta_v)$ depending on the activator concentration $\theta_c$ in it, which reflects the vascular network response. Here the function $R_n(\theta)$ is assumed to the identical for all the vein of one level $n$.  Second, to every branching point $k$ we ascribe a certain blood pressure $P$. Then the blood flow distribution is governed by the system of equations including Ohm's law written for every vein $v$ (Fig.~\ref{Fig10})
\begin{equation}\label{3.1}
    P_\text{in} - P_\text{out}  = J_v R_n(\theta_v)\,,
\end{equation}
and conservation of blood and activator substance at every branching point $k$
\begin{eqnarray}
\label{3.2}
    J_{\text{in},1} + J_{\text{in},2}  &=& J_{\text{out}}\,,
\\
\label{3.3}
    J_{\text{in},1}\theta_{\text{in},1} + J_{\text{in},2}\theta_{\text{in},2}
      &=& J_{\text{out}}\theta_{\text{out}}\,.
\end{eqnarray}
This system of equations should be completed by the ``boundary'' conditions specifying the blood pressure at the entrance to the venules $\{v_c\}$ and the exit from the root vein
\begin{equation}\label{3.4}
    P_{\text{in}|v_c}= P\,,\quad P_{\text{out}|n =0}=0
\end{equation}
and the equality of the activator concentration in every venule $v_c$ to the concentration of activators, or more rigorously, its mean value in the corresponding elementary domain $\mathcal{Q}_{v_c}$
\begin{equation}\label{3.5}
    \theta_{v_c}= \theta_{\mathcal{Q}_{v_c}}\,.
\end{equation}
The latter equality takes into account that blood flow through capillaries is rather slow and the capillary walls have numerous minute capillary pores permeable to water and other small molecular substances. This gives rise to local equilibrium between blood and the surrounding cellular tissue with respect to redistribution of biochemical compounds and makes the activator concentration a practically uniform field on scales about the size of elementary domain.

In the problem under consideration the distribution of activators is treated as a given beforehand field $\theta(\mathbf{r})$.  Solving the system of equations~\eqref{3.1}--\eqref{3.3} subject to conditions~\eqref{3.4} and \eqref{3.5} we can find the blood flow rate in the venules $\{v_c\}$. The latter together with expression~\eqref{eta} gives us the desired dependence of the blood perfusion rate $\eta\{\theta\}$ on the field $\theta(\mathbf{r})$.

\subsection{Regularities of the vessel response}

To complete the developed model we should specify the dependence of the hydrodynamic resistance $R_v(\theta_v)$ of every  vein $v$ on the activator concentration $\theta$ in it. According to the adopted assumptions the value $\theta_c =1$ ($\Theta_c = \Delta$ in dimensional units) characterizes the critical conditions when one of the homeostasis parameters comes close to the boundary of tolerance zone. To survive the living tissue has to increase essentially the blood perfusion rate which is implemented via vessel dilation. Since all the arteries of a microcirculatory bed contribute equally to its vascular resistance each artery should exhaust its ability to widen when the activator concentration exceeds the critical value $\theta_c = 1$. This allows us to adopt the following ansatz
\begin{equation}\label{3.7}
    R_v(\theta_v) = \rho_n\phi(\theta_v)\,,
\end{equation}
where $\rho_n$ is the vessel resistance at $\theta_v = 0$ which depends only the number of the hierarchy level and $\phi(\theta)$ is a certain function universal for all the vessel under consideration. Naturally, by definition, at $\theta = 0$ the equality $\phi(0) =1$ holds and for $\theta\gg1$ the inequality $\phi(\theta) \to \phi_\text{lim}\ll1$ should be the case.

\begin{figure}
\begin{center}
\includegraphics[width=60mm]{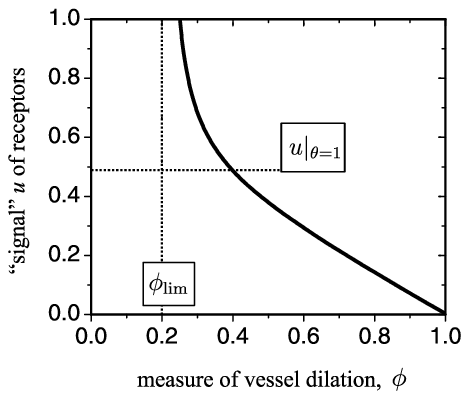}
\end{center}
\caption{The expected dependence between the vessel dilation measured in units of $\phi:=R/R_{\theta = 0}$ and the receptor signal $u(\theta)$ caused by activators of concentration $\theta$.}
\label{Fig11}
\end{figure}

\begin{figure}
\begin{center}
\includegraphics[width=70mm]{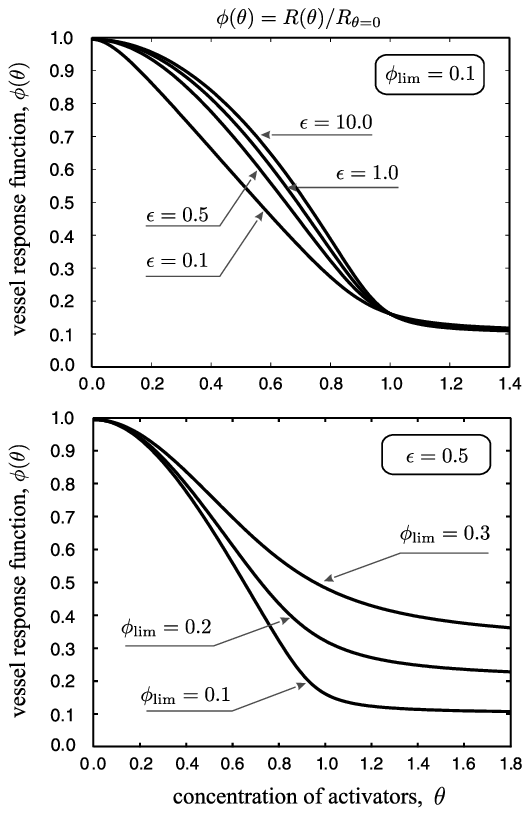}
\end{center}
\caption{The vessel response function $\phi(\theta)$ specified by ansatz~\eqref{3.10} for various values of its parameters. The upper fragment exhibits this function for $\phi_\text{lim} = 0.1$ and $\epsilon = 0.5$, 1.0, 10.0., the lower one does it for $\phi_\text{lim} = 0.1$, 0.2, 0.3 when the fixed value of $\epsilon = 0.5$.}
\label{Fig12}
\end{figure}

In order to construct the function $\phi(\theta)$ let us apply to the following speculations. There should be some threshold $\epsilon\ll1$ in the receptor ability to detect activators, meaning that activators cannot be detected if their concentration $\theta\lesssim\epsilon$. So, to be specific, we suppose that the ``signal'' $u(\theta)$ generated by the receptors is related to the activator concentration by the ansatz
\begin{equation}\label{3.8}
   u(\theta) = \sqrt{\theta^2 + \epsilon^2} - \epsilon\,.
\end{equation}
Then the expected relationship between the function $\phi(\theta)$ quantifying the vessel response and the receptor ``signal'' $u(\theta)$ should be of the form shown in Fig.~\ref{Fig11}. In particular, as $u\to\infty$ the value  $\phi\to\phi_\text{lim}$ and at $u = u(\theta =1)$ the vessel dilation should get such a degree that the value $\phi$ be about its limit; we set $\phi_{\theta=1} = 2\phi_\text{lim}$. The following ansatz
\begin{equation}\label{3.9}
  \frac{1-\phi_\text{lim}- 2\phi_\text{lim}^{2}}{1-\phi_\text{lim}}\cdot
  \frac{u_{\theta=1}-u}{u_{\theta=1}}
  =\phi-\frac{2\phi_\text{lim}^{2}}{\phi-\phi_\text{lim}}\,,
\end{equation}
has been chosen to specify this dependence. The combination of expressions~\eqref{3.8} and \eqref{3.9} yields the final ansatz for the vessel response function $\phi(\theta)$
\begin{align}
\label{3.10}
    \phi(\theta)& =\frac{\phi_\text{lim}+U(\theta)}{2}+\sqrt{\frac{[\phi_\text{lim}-U(\theta)]^{2}}{4}+2\phi_\text{lim}^{2}}
    \,,
\\
\intertext{where}
\label{3.11}
    U(\theta)&=\frac{1-\phi_\text{lim}-2\phi_\text{lim}^{2}}{1-\phi_\text{lim}}\cdot \frac{\sqrt{\epsilon^{2}+1}-\sqrt{\epsilon^{2}+\theta^{2}}}{\sqrt{\epsilon^{2}+1}-\epsilon}\,.
\end{align}
Figure~\ref{Fig12} visualizes the constructed function $\phi(\theta)$ for various values of its parameters, including $\epsilon = 0$ and $\phi_\text{lim} = 0$ matching the ideal vessel response. In some sense, ansatz~\eqref{3.10} can be justified applying directly to Fig.~\ref{Fig12} because the purpose of its construction is only to have the ability to analyze different cases of the living tissue response varying the parameters of an appropriate ansatz.

The dependence of the vessel resistance on the hierarchy level, i.e. the value of $\rho_n$ vs the level number $n$ is constructed using the statement about the equality of contributions to the vascular resistance of arterial beds (Sec.~\ref{PBack}) from all the hierarchy levels. In other words, when the activator concentration is the same at all the points of the cellular tissue, in particular, $\theta(\mathbf{r}) = 0$, and, thus, all the vessels of one level have the same hydrodynamic resistance, the blood pressure drop should be uniformly distributed over the vascular network. This enables us to write the $\rho_n$-dependence in the form
\begin{equation}\label{3.12}
    \rho_n=(2\zeta)^{n}\rho_0\,,
\end{equation}
where the value $\rho_0$ is related with the root vein and the parameter $\zeta\approx1$ allows us to consider the cases where various groups of vessels dominate in the blood pressure distribution over the venous bed. In particular, for $\zeta < 1$ the large veins contribute mainly to the vascular resistance of the vein bed, whereas, in the opposite case, $\zeta > 1$, the small veins are dominating.

\subsection{Numerical algorithm}

\begin{figure}
\begin{center}
\includegraphics[width=65mm]{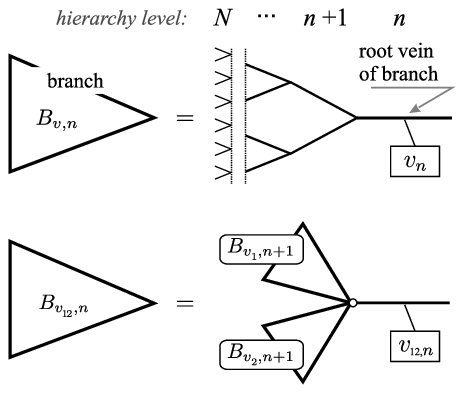}
\end{center}
\caption{The structure of a vein branch $B_{v,n}$ having the vein $v_n$ as a root vessel (upper line) and the interconnection between the branches of neighboring levels (lower line).}
\label{Fig13}
\end{figure}

The developed model was analyzed numerically. So before presenting the results the numerical algorithm used in solving equations~\eqref{3.1}--\eqref{3.3} is briefly described. In addition to veins we consider the collection of vein branches $\{B_v\}$, each of which originates from one $v$ of the veins and contains all its generations (Fig.~\ref{Fig13}). The hierarchy of branches stems directly from the hierarchy of veins by ordering the branches according to the position of their root vessels in the vein tree. To every branch $B_{v}$ we ascribe the resistance to blood flow through it
\begin{equation}\label{new:1}
    Z_{v} = \frac{P_{\text{in},v}}{J_{v}}
\end{equation}
just dividing the blood pressure drop across the branch, i.e. $P_{\text{in},v}$ (see Fig.~\ref{Fig10}) by the blood flow rate $J_v$ at its root vein $v_{n}$. To every branch $B_{v}$ we also ascribe the activator concentration $\theta_v$ in blood flow going through its root vein $v$.

For the venules (veins of the last level) their branches $\{B_{v_c}\}$ consist only of these vessels. So their resistances $\{Z_{v_c} = \rho_N\phi(\theta_{v_c})\}$ can be found directly by virtue of condition~\eqref{3.5}. To pass to the preceding level $N-1$ we take into account the interconnection between the branches of the neighboring levels shown in Fig.~\ref{Fig13}.  Writing Ohm's law for these branches (see Exp.~\eqref{new:1}) and taking into account mass conservation at the branching point (see Exps.~\eqref{3.2}, \eqref{3.3}) we, at the first step, can find the activator concentration $\theta_{12,n}$ in the vein $v_{12,n}$ from the formula
\begin{equation}\label{4.1}
\theta_{12,n}  = \left[ {\frac{{\theta _{1,n + 1} }}
{{Z_{1,n + 1} }} + \frac{{\theta _{2,n + 1} }}
{{Z_{2,n + 1} }}} \right]\left[ {\frac{1}
{{Z_{1,n + 1} }} + \frac{1}
{{Z_{2,n + 1} }}} \right]^{ - 1}
\end{equation}
if the values $\{\theta_{n+1}\}$ of activator concentration and the branch resistances $\{Z_{n+1}\}$ are know at level $(n+1)$, in particular, we know the quantities $\theta_{1,n+1}$, $\theta_{2,n+1}$, $Z_{1,n+1}$, and $Z_{2,n+1}$ ascribed to the branches $B_{v_1,n+1}$ and $B_{v_2,n+1}$.
Then, at the next step, we calculate the resistance $Z_{12,n}$ of the branch $B_{v_{12},n}$ from the equality
\begin{equation}\label{4.2}
\frac{1}
{{Z_{12,n}  - R_n (\theta _{12,n} )}} = \frac{1}
{{Z_{1,n + 1} }} + \frac{1}
{{Z_{2,n + 1} }}\,.
\end{equation}
In this way all the characteristics of level $(N-1)$ can be found using these data of level $N$ and so on up to the root vein of level 0. After reaching the root vein the total blood flow rate $J_0 = P/Z_0$ is found. Then going in the opposite direction we calculate the blood flow rates $\{J_1\}$ at level 1 using the equality
\begin{equation}\label{4.3}
J_{1(2),n + 1}  = J_{12,n} \frac{1}
{{Z_{1(2),n + 1} }}\left[ {\frac{1}
{{Z_{1,n + 1} }} + \frac{1}
{{Z_{2,n + 1} }}} \right]^{ - 1}
\end{equation}
describing the blood flow split at branching points and so on until getting the venules. At the final step expression~\eqref{eta} gives us the desired dependence of the blood perfusion rate on the distribution of activators in cellular tissue.

\section{Results of simulation}

The blood flow redistribution governed by equations~\eqref{3.1}--\eqref{3.3} subject to conditions~\eqref{3.4} and \eqref{3.5} was studied numerically using the algorithm described in the previous section. Variations in the blood flow rate at every vein $v$ as well as in the blood perfusion rate are analyzed in units of the corresponding quantities, $I_{n}$ or $\eta_0$, matching the absence of the activators, $\theta(\mathbf{r}) = 0$. These units allow us to reduce the number of the system parameters; only the total number of levels $N$, the parameter $\zeta$ measuring the relative contribution from the lager and small vessels to the vascular network resistance, and the parameters $\epsilon$ and $\phi_\text{lim}$ of the individual vessel response are essential.

\begin{figure*}
\begin{center}
\includegraphics[width=125mm]{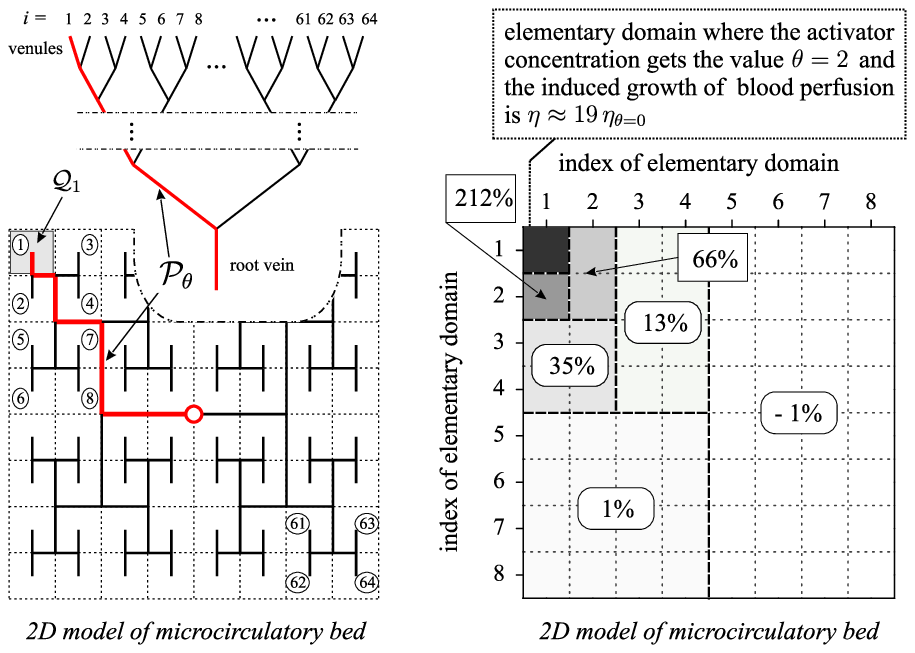}
\end{center}
\caption{(Color online) Example of the systems analyzed numerically. The region with a nonzero value of the activator concentration $\theta$ is shadowed and the ordering of the vanules is illustrated for the system with the number of the hierarchy levels equal to $N=6$. The right fragment visualizes the blood flow distribution in the domain of microcirculation bed based on the data obtained numerically and presented in Fig.~\ref{Fig15}.}
\label{Fig14}
\end{figure*}

The given section presents the results for the 2D system when the activator concentration $\theta(\mathbf{r})$ differs from zero inside $m$ first elementary domains located at one of the square corner, namely,
\begin{equation}\label{5.1}
\theta ({\mathbf{r}}) =
\begin{cases}
  \theta, & {\mathbf{r}} \in   \mathcal{Q}_\theta := \bigcup_{1\leq i \leq m} \mathcal{Q}_i   \\
  0, &     {\mathbf{r}} \notin \mathcal{Q}_\theta \,.
\end{cases}
\end{equation}
Figure~\ref{Fig14} illustrates the analyzed case and shows the introduced ordering of the venules with index $i$. The elementary domains are also labeled with the index $i$ of the corresponding venule.

\begin{figure*}
\begin{center}
\includegraphics[width=154mm]{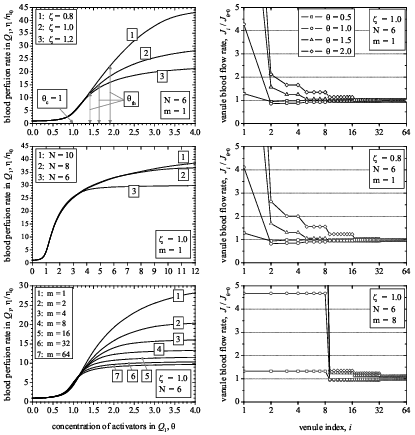}
\end{center}
\caption{Results of numerical simulation. Blood perfusion rate $\eta$ (in units of $\eta_0$) vs the concentration of activators in the same elementary domain $\mathcal{Q}_1$ (left column) and the distribution of the blood flow rates (in units of $J_{i,N,\theta = 0}$) over the venules, i.e. over the vessels of the last hierarchy level (right column). In obtaining these results the parameters $\epsilon = 1.0$ and $\phi_\text{lim} = 0.1$ were used, the used values of the other parameters are shown in figure.}
\label{Fig15}
\end{figure*}

Figure~\ref{Fig15} presents the obtained results of numerical simulation. The first column exhibits the dependence of the blood perfusion rate $\eta(\theta)$ on the activator concentration $\theta$ at the same point (through the same elementary domain) of living tissue. In obtaining these results activators were assumed to be located only in the first elementary domain $\mathcal{Q}_1$ ($m=1$, except for the case shown in the left bottom frame) and the number of the hierarchy levels was set equal to $N=6$ (but the case shown in the left middle frame). It meets tenfold decrease in the vessels length when passing from the root vessel to the smallest arterioles or venules.

The left upper frame depicts three curves corresponding to different values of the parameter $\zeta$. When the activator concentration in the domain $\mathcal{Q}_1$ gets the threshold $\theta_c = 1$ of the individual vessel response the venule joined with this domain  exhausts its capacity to widen. However it is not enough for the blood flow rate to grow substantially. Indeed, the contribution of this venule to the vascular resistance is small even along the path $\mathcal{P}_\theta$ leading from the root vessel to the domain $\mathcal{Q}_1$ on the vein tree (Fig.~\ref{Fig14}). So many veins along this path have to dilate in order for the blood perfusion rate to grow essentially in $\mathcal{Q}_1$. However, the larger a vein, the larger the tissue region drained through this vein as a whole and its size increases exponentially with the number of branching points separating the venule $v_1$ and the given vein along $\mathcal{P}_\theta$. As a result when the activator concentration in the domain $\mathcal{Q}_1$ is about $\theta_c = 1$ the activator concentration in relatively large veins on the path $\mathcal{P}_\theta$ turns out to be rather small and these vessels cannot widen remarkably. Thereby the blood flow rate cannot also exhibit substantial increase (Fig.~\ref{Fig15}). As the activator concentration grows further the induced increase in the activator concentration along the path $\mathcal{P}_\theta$ gets the critical value $\theta_c = 1$ in larger and larger veins, causing the growth of the blood perfusion rate in the domain $\mathcal{Q}_1$. When the main part of the veins belonging to $\mathcal{P}_\theta$ widen to the maximum the blood perfusion rate comes to the upper limit $\eta_\text{max}$.

The maximum $\eta_\text{max}$ of the blood perfusion rate depends on the system parameter, in particular, on $\zeta$. As it will be demonstrated below practically all the additional amount of blood entering the system due to the vessel dilation is directed to the domain $\mathcal{Q}_1$. So the higher the contribution of large veins, the higher the upper limit of the blood perfusion rate. These speculations are justified well in Fig.~\ref{Fig15}, the left upper frame. Because of this blood flow focusing along the path $\mathcal{P}_\theta$ the upper limit $\eta_\text{max}$ of the blood perfusion rate can exceed substantially the value $1/\phi_\text{lim}$ which would be attained if the total blood pressure drop had fallen just on the last vessel supplying the region $\mathcal{Q}_1$ with blood.

A similar effect should be expected for the dependence of the blood perfusion rate on the number of hierarchy levels. The higher this number, the larger the upper limit $\eta_\text{max}$  of the blood perfusion rate, which is demonstrated directly in Fig.~\ref{Fig15}, the left middle frame. When the region of living tissue where  activators are located becomes larger it effectively renormalizes the size of elementary domains and decreases the number of hierarchy levels. So the value of $\eta_\text{max}$ has to drop with $m$ increasing. The latter behavior is demonstrated in Fig.~\ref{Fig15} (left bottom frame), in particular, the value $\eta_\text{max}$ goes down to $\phi_\text{lim}$ when activators spread over the whole region of the microcirculatory bed.

The obtained results demonstrate us also the fact that when the vessel behavior is not ideal the threshold $\theta_\text{th}$ in the vascular network response as a whole does not coincide with the critical value $\theta_c=1$ of the individual vessel reaction and can exceed it remarkably. Let us define the threshold $\theta_\text{th}$ as the value at which the blood perfusion rate attains one half of its upper limit. Than this fact is clearly visible in Fig.~\ref{Fig15} (left upper frame) and, for example, the estimate $\theta_\text{th}\approx 2$ holds for the case shown by curve~1. Naturally, the threshold $\theta_\text{th}$ depends also on the parameters of the vascular network but not only on the characteristics of individual vessel behavior.

The shown curves enable us to assume that due to the fractal structure of the vascular network there is a certain universal function $\eta(\theta|\phi,\epsilon)$ depending only on the parameters of individual vessel behavior, which describes the increase in the blood perfusion rate with the growth of the activator concentration until the latter becomes greater then the threshold $\theta_\text{th}$. The other characteristics of the vascular network, roughly speaking, affect only the upper limit $\eta_\text{max}$ of blood perfusion rate.

The right column depicts the distribution of the blood flow rates over the venules. Comparing the data shown hear and that of the left column we can declare that the main increase in the blood perfusion rate is located in the region of living tissue containing activators. Exactly these data enables us to assumed that the additional amount of blood flow has to be directed mainly along the path $\mathcal{P}_\theta$. However, the question about the interference of blood streams induced by distant regions of living tissue requires an individual investigation. Nevertheless, when the living tissue is excited only in one domain its response can treated as quasilocal, at least at the first approximation. We have used the quasilocal term because this response depends also on the size of the excited domain.

As the nonlocal component of the $\eta\{\theta\}$-functional is concerned, the right column of Fig.~\ref{Fig15} demonstrates us that depending on the relation between the activator concentration $\theta$ in the elementary domain $\mathcal{Q}_1$ and the vascular network threshold $\theta_\text{th}$ it changes its form. When $\theta< \theta_\text{th}$ the increase in the blood flow rate through $\mathcal{Q}_1$ is caused by the blood flow decrease through the other elementary domains. In this case the required growth of the blood perfusion rate is due to  the redistributed of blood flow over the vascular network without additional amount of blood entering the microcirculatory bed. Since the number of the surrounding domains is rather large the decrease in the blood perfusion rate outside the excited region is not essential. When $\theta> \theta_\text{th}$ the increase in the blood perfusion rate is mainly caused by additional portion of blood entering the microcirculatory bed. Therefore the blood perfusion rate is increased in the surrounding domains also.

The resulting distribution of the blood perfusion rate in living tissue is shown in Fig.~\ref{Fig14} (right fragment) for the 2D model. As seen the distribution can be rather heterogeneous because the elementary domains neighboring in the physical space can be supplied with blood through the veins (and arterioles) belonging to substantially different branches. If activators are located in another elementary domain the found distribution will also describe this case after a simple reordering of the vessels whereas in the physical space the blood perfusion pattern can become more heterogeneous then the shown one. The heterogeneity of the blood perfusion rate is caused by the presence of two topologies in the system, one is related to physical space and the other corresponds to the vascular network architectonics.

\section{Conclusion}

The paper develops a model for the living tissue as an example of natural complex systems with active behavior. The necessity for such systems to keep their parameters within tolerance zone makes their self-regulation crucial. Many of these systems like living tissue and consumer goods markets are organized hierarchically, i.e. have highly branching networks supplying their elements with the required ``nutrients''. The absence of central governing units poses a question about the mechanism of self-regulation by which the ``nutrients'' are delivered to the points where it is necessary and the equilibrium at the other points is not disturbed. When an element requires additional amount of ``nutrients'' the supply network responding at all the levels should deliver them.

Keeping in mind the previous papers devoted to this problem \cite{we1,we2,we3,we4} we have developed a model for a microcirculation bed (a unit of regional circulation) embedded in 2D and 3D dimensional domains of cellular tissue in the self-filling manner. The cells of the tissue continuously require oxygen, various nutrient substances and the products of their live activity, for example, carbon dioxide, should be withdrawn. When the intensity of cell activity grows the supply of these compounds has to be increased. Exactly this function of the vascular network is under investigation. We have assumed that the necessity for the increased blood supply is signalized by cells via the release of some activators, special biochemical compound.

The artery and vein networks are considered to be of the tree form made up of the artery-vein counter-current pairs. Activators produced by cellular tissue are withdrawn by flood flow leaving the system via the venous bed. The considered mechanism of living tissue self-regulation is based on two effects. The first one is conservation of blood flow and the amount of activators at the branching points of the vein network. It gives rise to the information self-processing which is necessary for vessels to respond individually in an adequate way. The hydrodynamic resistance of arteries is determined by the concentration of activators in the blood flow going through the vein forming with a given artery a counter-current pair. For the sake of simplicity this model is reduced to the description of only the venous bed. The dependence of the vessel resistance on the activator concentration is constructed in a rather general way and deviates substantially from the ideal form analyzed previously \cite{we1,we2,we3}.

The following results have been obtained. First, it is shown that the threshold $\Theta_\text{th}$ in the vascular network response to variations in the activator concentration as a whole can differ essentially from the critical value $\Delta$ of the individual vessel response and is determined not only by the parameters of the vessel behavior but also the characteristics of the vascular network architectonics.

Second, due the fractal structure of the vascular network the dependence of the blood perfusion rate on the activator concentration in the cellular tissue is quasilocal, at least, at the first approximation. This means that the blood perfusion rate $\eta(\Theta)$ depends mainly by the activator concentration $\Theta$ taken at the same point of the living tissue. The prefix ``quasi'' takes into account that this dependence is affected also by the size of the excited living tissue region. The results of numerical simulation have demonstrated the fact that the $\eta(\Theta)$-dependence is of a certain universal form determined only by the properties of the individual vessel response for not too high values of the activator concentration, $\Theta \lesssim \Delta$ and only the upper limit  $\eta_\text{max}$ of the blood perfusion rate is affected by the characteristics of the vascular network architectonics and the size of the excited region. It should be reminded that the blood perfusion rate under the normal conditions is treated as a known value. This found quasilocality is explained by the fact that the main part of additional amount of blood going through the system is focused along a path on the vascular network connecting the excited region and the root vessels.

Third, due to the presence of two topologies in living tissue, the point proximity in the physical space and the vessel proximity with respect to their position in the vascular network the nonlocal component of the functional  $\eta\{\Theta\}$-dependence can be rather heterogeneous in space because neighboring points of living tissue can be supplied with blood by vessels belonging to vessel branches distant from each other in the vascular network architectonics.

\acknowledgments

The authors appreciates the support of DFG Grant MA 1508/8-1 as well as RFBR Grants 06-01-04005 and 09-01-00736.

\end{document}